\documentclass{article} 

\usepackage{url,hyperref,graphicx,textcomp}
\def\si2r2{SI$_2$R$_2$}

\usepackage{color}

\begin{document}

\title{The hidden side of COVID-19 spread in Italy}

\author{Luigi Brugnano\,\footnote{Dipartimento di Matematica e Informatica ``U.\,Dini'', Universit\`a di Firenze, 50134 Firenze, Italy.
         Tel.: +39\,055\,2751421, e-mail:\,{\tt luigi.brugnano@unifi.it}}  \and Felice Iavernaro\,\footnote{Dipartimento di Matematica, Universit\`a di Bari, 70125 Bari, Italy.
 E-mail:\,{\tt felice.iavernaro@uniba.it} } \and Paolo Zanzottera\,\footnote{Dipartimento di Economia e Management, Universit\`a di Brescia, 25122 Brescia, Italy. E-mail:\,{\tt paolo.zanzottera@unibs.it}}
}

\maketitle


\begin{abstract} 
\noindent{\bf Background.} The paper concerns the SARS-CoV2 (COVID-19) pandemic that, starting from the end of February 2020, began spreading along the Italian peninsula, by first attacking small communities in north regions, and then extending to the center and south of Italy, including the two main islands. 

\medskip
\noindent{\bf Objective.}  The creation of a forecast model that manages to alert the decision-making bodies and, in particular, the healthcare system, to hinder the emergence of any other pandemic outbreaks, or the arrival of subsequent pandemic waves. 

\medskip
\noindent{\bf Methods.} A new mathematical model to describe the pandemic is given. The model includes the class of undiagnosed infected people, and has a multi-region extension, to cope with the in-time and in-space heterogeneity of the epidemic.

\medskip
\noindent{\bf Results.} We obtain a robust and reliable tool for the forecast of the total and active cases, which can be also used to simulate different scenarios.

\medskip
\noindent{\bf Conclusions.} We are able to address a number of issues, such as assessing the adoption of the {\em lockdown} in Italy, started from 11 March 2020, and how to employ a rapid screening test campaign for containing the epidemic.

\bigskip


\end{abstract}

\newpage
\section{Introduction}\label{intro}

Virus SARS-CoV2 hit Italy at the end of February 2020: after the first patient diagnosed in Codogno (Lodi, Lombardy) on the 21st of February, the virus spread very quickly in the Italian peninsula, particularly in Lombardy and in the northern Italy. The syndrome of acute respiratory disease and the consequences of COVID-19 put the national healthcare system under stress.
The first epidemic phase was characterized by a huge number of patients with severe or critical conditions that congested Lombardy hospitals, particularly in the provinces of Lodi, Bergamo, and Brescia. Also neighboring provinces of the Emilia Romagna region were affected: hospitals of provinces of Piacenza and Parma had great difficulties to manage COVID-19 patients. Disease spread in the regions around the epicenter of Lodi, but Italian government closed the country progressively: at the beginning a red zone around Lodi was created; then, all the Lombardy was declared red zone; finally, on 11 March 2020 every region was locked down.\footnote{In more details,  starting from 10 March it was forbidden to move from one region to another one, and every non essential activity was suspended on the 11th of March.}
 In Italy, the COVID-19 spread has been inhomogeneous: in fact, in the North the virus spread out before the lockdown and a huge number of people contracted the virus, whereas the Centre, South and Italian islands regions (Sicily and Sardinia) had fewer cases, so that these regions were able to manage the disease. This regional differences were amplified by the lockdown in the country.

\begin{figure}[t]
\centerline{\includegraphics[width=8cm]{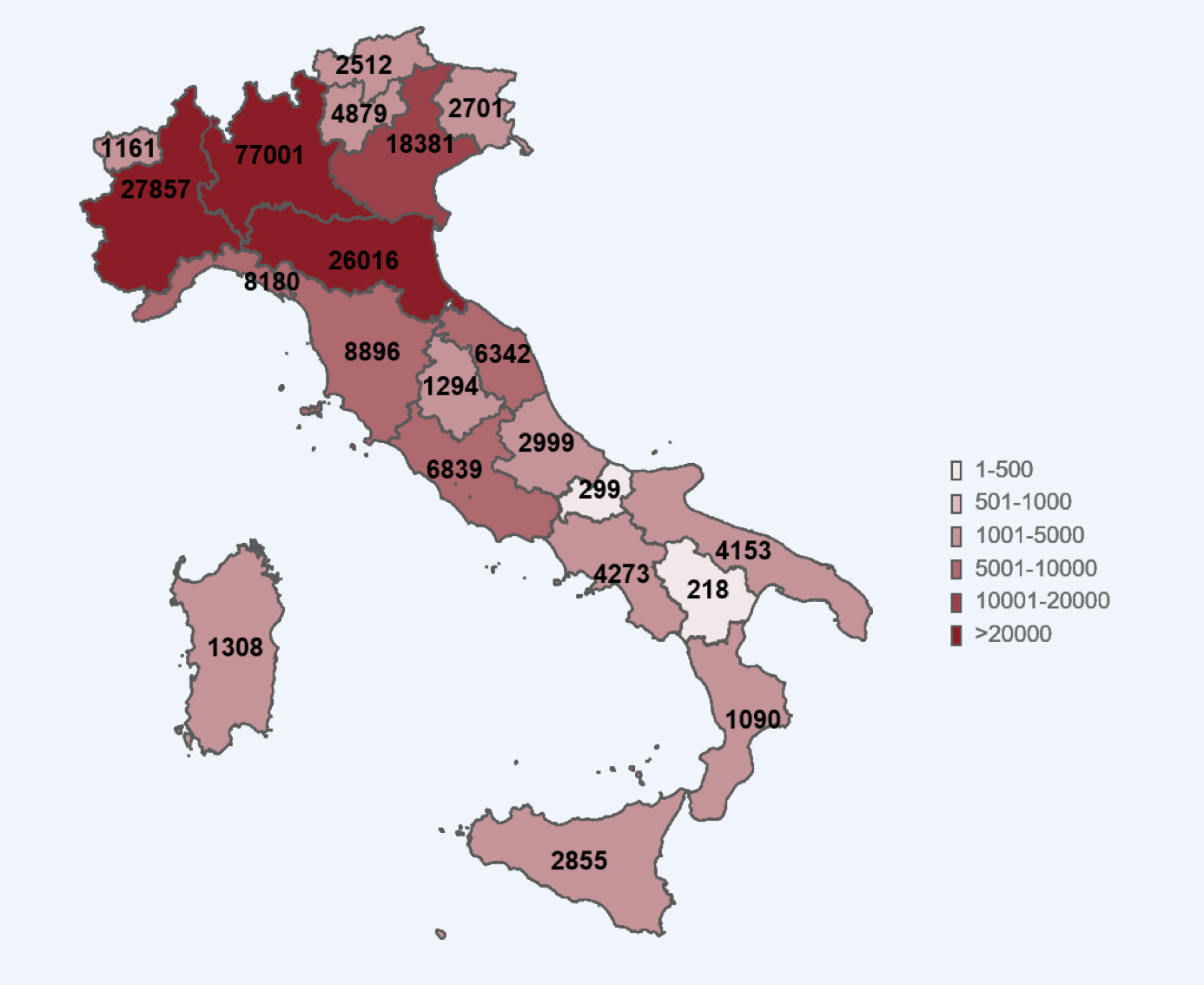}}
\caption{COVID-19 cases in the different regions of Italy on 4 May 2020 (source Italian Istituto Superiore di Sanit\`a).}
\label{fig_ISS1}
\end{figure}

An inhomogeneous spread of COVID-19 happened also in other countries and, where it coincides with a high-density residential area and with high mobility, the results are dramatic, as it happened in Lombardy and in the metropolitan area of big cities, such as Madrid, London, and New York. In those cases it is fundamental the capacity of resilience and resistance of the health sanitary system in the region where the spread exploded as a big pandemic wave. For these reasons mr\si2r2, the model given here, is a multi-regional model that takes into account the inhomogeneous spread of the disease.
Figure~\ref{fig_ISS1} shows how there is a huge regional difference in the SARS-CoV2 spread along the country.

\begin{figure}[t]
\centerline{\includegraphics[width=11.8cm]{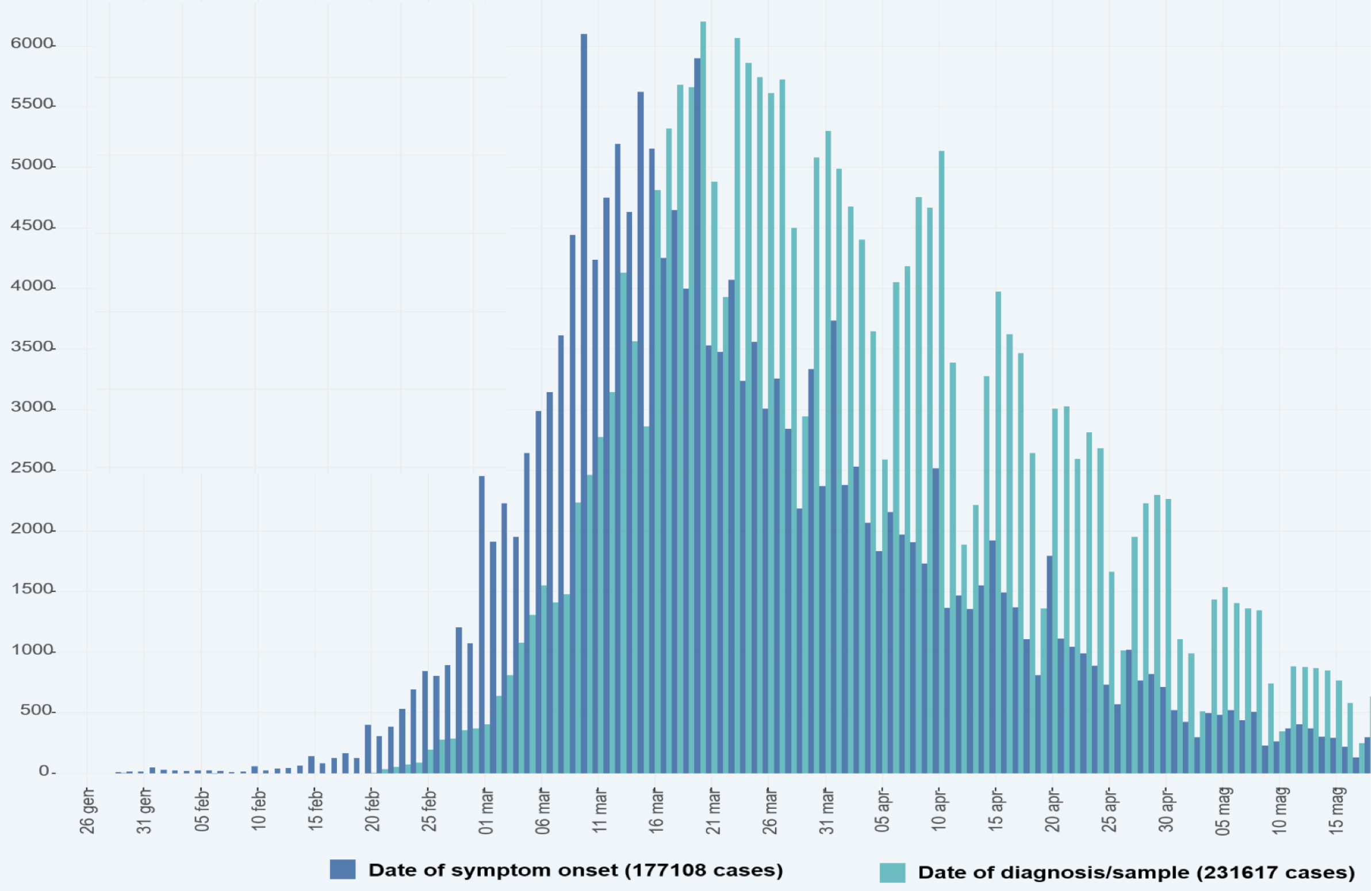}}
\caption{diagnoses and onset of symptoms for patients in Italy (source Italian Istituto Superiore di Sanit\`a) \cite{delay}.}
\label{fig_ISS2}
\end{figure}

The incubation period of COVID-19 is quite long: for this reason when the first patient, not imported from China, was diagnosed in Lombardy on the 21st of February, SARS-CoV2 was already spread in the most populated Italian region. The virus was circulating since the beginning of February in Italy, as is demonstrated by the onset data released by the Health National Institute. Once people are diagnosed positive to COVID-19 from tests,  they are asked  when they felt first symptoms and the onset data are drawn on the graph (see Figure~\ref{fig_ISS2} ). In fact, we never talked in Italy about the patient-zero, but it was always considered at least the patient-one. 

For this reason the Italian healthcare system, as other European national health systems in Spain, France, UK, and Belgium, were not prepared to manage the first epidemic wave in Europe. The course of the COVID-19 disease is quite long: it takes several days from the virus contact to the first symptoms to hospitalization to death or recovery. Fastest cases take more than one week and many other cases have a disease course of several weeks of hospitalization.

It is impossible to test the entire population of a country in a short time, to verify the presence of SARS-CoV2 virus. Above all, at the beginning of the epidemic, tests were scarce. A big part of the population has contracted the virus, but it is not included in the ``positive cases'' statistics: both for the presence of asymptomatic people, and for the difficulty of testing people with clear COVID-19 symptoms. A healthcare system has to be prepared in the future to test hundreds of thousands people each day. Tests are really important to fight the virus as they are able to seek positive people, to isolate them, and to contain the virus spread.

The lack of ICU beds, medical devices such as lung fans and even oxygen, have amplified the impact of COVID-19 in the most affected areas. To this must be added the lack of an adequate number of doctors, nurses and health workers and, in some cases, even that of personal protective equipment. All this, combined with the fact that COVID-19 is a new disease, little known even in the medical field, has increased the apparent lethality rate, particularly in the areas most affected by the spread of the virus.
The preparation of the national healthcare system is therefore decisive both for the containment of the disease and for its ability to limit its most harmful effects on people. The creation of a forecast model that manages to alert the decision-making bodies and, in particular, the healthcare system, is therefore decisive to hinder the emergence of any other pandemic outbreaks, or the arrival of subsequent pandemic waves. 

The mathematical model mr\si2r2 has been introduced for this reason, and has proved to be able to predict the trend of the positives cases with an accuracy greater than 90\%, 3 weeks in advance. Having more than 20 days to prepare for the management of a further pandemic outbreak, even on a regional and local basis, is essential to set up suitable containment measures in due time.  
Based on an extension of the classical SIR equations (see, e.g., \cite{Lu79}), the new model takes the form of a delay differential system which incorporates the class of \emph{undiagnosed infected individuals}: these are the actual responsible for virus shedding during their infectivity period, unless they are detected for some reason and, therefore, quarantined. Within this class are asymptomatic and pauci-symptomatic people, as well as infectious pre-symptomatic people, considering that a significant portion of transmission is expected to occur before infected persons have developed symptoms \cite{Xi2020}. To take account for the time lag between infectiousness onset, appearance of symptoms, and infection detection, a delay time $\tau$ is introduced in the equations, making the resulting system very well suited to fit the observed data. Furthermore, to account for the in-space and in-time heterogeneity of the epidemic diffusion, the Italian peninsula has been subdivided into four macro-areas and a new term has been introduced  in the equations to consider the effect of   a possible migration from one area to another, given that even a small uncontrolled flow of infected people may cause the appearance of a new source of infection in a clean zone.      

The mathematical modeling of the COVID-19 spread has been the subject of many investigations (see, e.g. \cite{FaPi2020,XiFu2020,Gior2020,ReRe2020}). The model here considered, though quite refined in the description of the underlying infection mechanisms, tries to keep at minimum the number of the involved variables and parameters to be estimated, in order to reduce the risk of redundancy (existence of multiple solutions) and, at the same time, to provide a reliable and computationally affordable forecast tool. Among the other features, this model is able to predict the actual portion of infected people (attack rate), to outline the benefits of the lockdown, and to assess the extent to which rapid screening tests may help in mitigating the risk of a future escalation of the infection.  

With these premises, the structure of the paper is as follows: in Section~\ref{mrsi2r2model} we present the \si2r2 model and its multi-region extension (mr\si2r2); in Section~\ref{results} we study a number of application of the model to the spread of the epidemic in Italy; finally, a few concluding remarks are reported in Section~\ref{fine}.

\section{Methods}\label{mrsi2r2model}
\begin{figure}[t]
\centerline{\includegraphics[width=9cm]{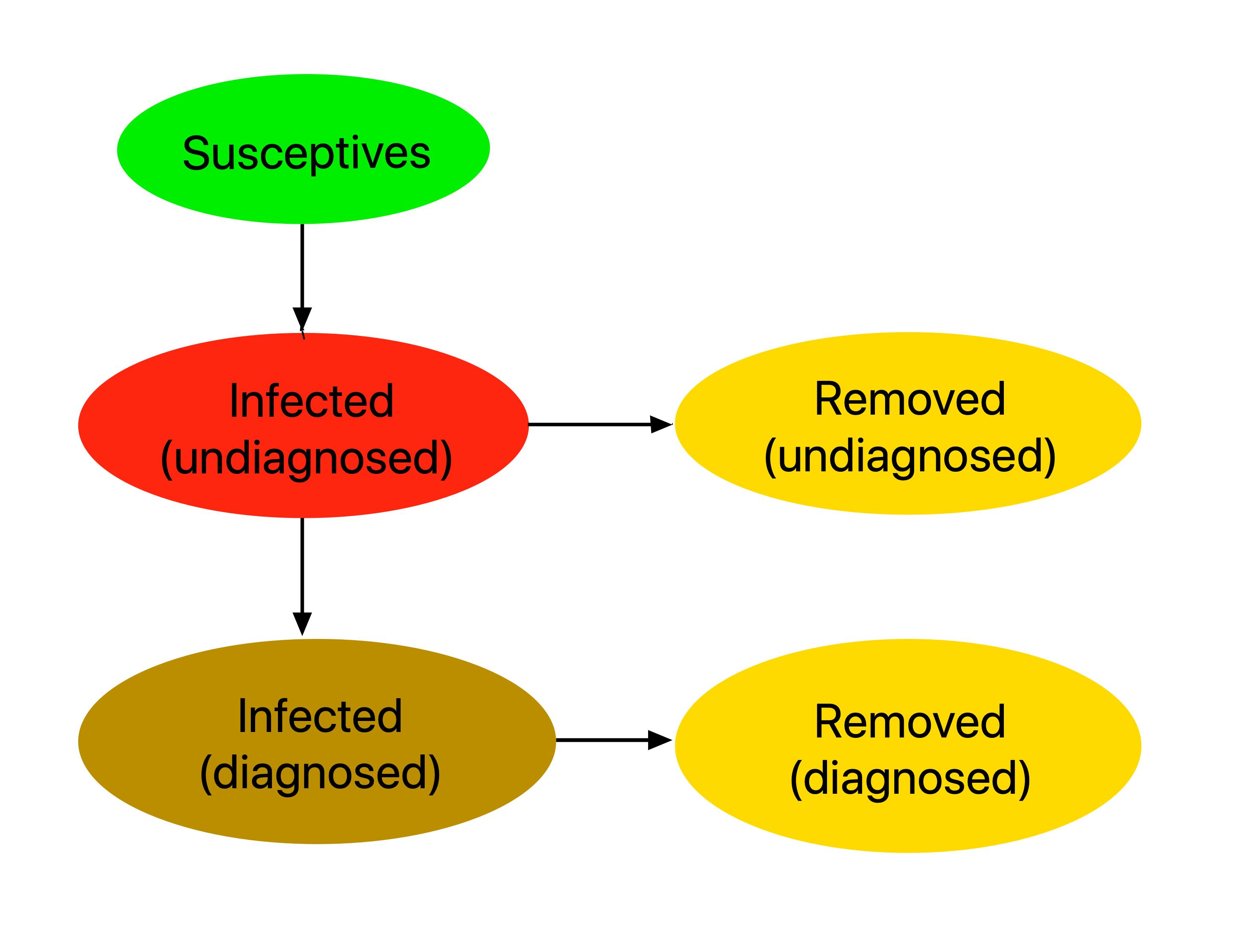}}
\caption{interactions among the classes of the \si2r2 model.}
\label{intera}
\end{figure}

In case of a single region, we divide the population into the following 5 classes:
\begin{itemize}
\item[$S$:] {\em Susceptive} individuals,  circulating healthy people who come in contact with infected persons and, thus, can catch the virus;
\item[$I_1$:] {\em Infected undiagnosed} individuals, the subclass of infected people who can spread the disease (carriers);
\item[$I_2$:]  {\em Infected diagnosed} individuals, the subclass of infected people who tested positive for COVID-19 and, thus, quarantined;
\item[$R_1$:] {\em Removed undiagnosed} individuals, spontaneously recovered people coming from class $I_1$;
\item[$R_2$:] {\em Removed diagnosed} individuals,  either recovered or died people coming from class $I_2$.
\end{itemize}
Recovering from the illness seems to give some immunity \cite{Bao2020,Lan2020}, so that recovered people are assumed to be not susceptive anymore. The diagram in Figure~\ref{intera} elucidates the interactions between the above classes defining the basic \si2r2 model.\footnote{The acronym \si2r2 derives from the initials of the 5 classes.}

Concerning our definition of susceptive people (class $S$) a remark is in order. In contrast with what is usually done in the literature, we leave the initial value of $S$ as an unknown parameter. Its value has to be determined  during the optimization procedure, which estimates the set of the involved parameters in order to fit the observed data as close as possible. We made this choice for two main reasons:

\begin{itemize}
\item on one hand, the  assumption underlying general epidemic models that the population is subject to a form of homogeneous mixing (see, e.g. \cite{Lu79}),  evidently does not apply in our context: as was outlined in the introduction, local sources of infection have spread the disease in neighboring areas first, so that the number of healthy people coming in contact with the virus was far from being equal to all the population during the first stages of the epidemic diffusion;

\item on the other hand, our choice reflects the final goal of social-distancing measures, such as the lockdown, where people are isolated to prevent possible contacts with the virus. In this regard, it should be noted that in some regions the spread has been confined to small areas and, thus, did not affect the whole population that much.

\end{itemize}

Hereafter, in order to keep the notation as simple as possible, we shall make the following formal correspondences between compartments and variables describing their numerosity:
$$
\begin{array}{rcl}
\mbox{\em Susceptive individuals $S$} &\rightarrow& x\\
\mbox{\em Infected undiagnosed individuals $I_1$} &\rightarrow& y_1\\
\mbox{\em Infected diagnosed individuals $I_2$ } &\rightarrow& y_2\\
\mbox{\em Removed undiagnosed people $R_1$} &\rightarrow& z_1\\
\mbox{\em Removed diagnosed people $R_2$} &\rightarrow& z_2\\
\end{array}
$$
The  \si2r2 model is then described by the following equations,  
\begin{equation}\label{si2r2}
\left\{ \begin{array}{rcl}
\dot x(t)     &=& -\frac{\beta}N x(t)y_1(t),\\[1mm]
\dot y_1(t) &=& \frac{\beta}N x(t)y_1(t)-\sigma y_1(t-\tau) s_+(y_1(t))-\gamma_1 y_1(t),\\[1mm]
\dot y_2(t) &=& \sigma y_1(t-\tau)s_+(y_1(t))-\gamma_2 y_2(t),\\[1mm]
\dot z_1(t) &=& \gamma_1y_1(t),\\[1mm]
\dot z_2(t) &=& \gamma_2y_2(t),
\end{array}\right.
\end{equation}
where:
\begin{itemize}
\item $\beta$ is the {\em coefficient of infection rate};
\item $\gamma_1$ and $\gamma_2$ are the {\em removal coefficients} of the two classes $y_1$ and $y_2$, respectively;
\item $\sigma$ is the {\em probability of detection of infected people} by means of the Covid tests campaign; 
\item $\tau$ is a {\em delay time}, and accounts for the time elapsing between the onset of infectiousness and the detection of the disease, for example after the appearance of  symptoms or a contact-tracing procedure; 
\item $s_+(y_1(t)) = \left\{\begin{array}{ccl} 1, &~&\mbox{if} \quad y_1(t)>1,\\[1mm] 0,&&\mbox{otherwise},\end{array}\right.$ \\ is a switch factor which prevents the corresponding solution component to became negative.
\end{itemize}

Considering that the sum of the right-hand sides in (\ref{si2r2}) identically vanishes, we get the conservation property
\begin{equation}\label{N}
x(t)+y_1(t)+y_2(t)+z_1(t)+z_2(t) \equiv N,
\end{equation}
the total number of people entering the model, which remains constant over time.

All  the parameters listed above were left free, with the exception of the removal coefficient $\gamma_1$ and the delay time $\tau$, which have instead been inferred from the literature. The remaining parameters have been tuned in order to assure a good fitting with the observed data by employing a global optimization procedure. The reason for fixing $\gamma_1$ a priori is that, otherwise, a certain parameter redundancy was experienced and, consequently, the model appeared not locally identifiable. This means that different but close configurations of the parameters could be selected leading to essentially  the same accuracy in fitting the observed data. Based on the results reported in \cite{Noh2020}, we have set
\begin{equation}\label{gamma1} 
\gamma_{1} = 4.3\cdot 10^{-2} \,d^{-1}\approx (23 \,d)^{-1},
\end{equation}
where $d$ stands for {\em days}.  As for the choice of the delay time $\tau$, we have confined the analysis to symptomatic individuals only, and considered two additive contributions: the period of infectiveness before the development of symptoms  plus the  the time duration from the onset of symptoms and diagnosis. Their mean values have been inferred from  \cite{Xi2020} and the data available at \cite{ISS} respectively, resulting in the choice $\tau=10\, d$. Figure~\ref{fig_ISS2} better clarifies this aspect. It shows bar diagrams on the dates of onset of symptoms (dark blue bars) and dates of diagnosis (light blue bars). We have noted that the peak of onset date was on the 10th of March, the last day before Italy tightened the lockdown. Positives on diagnoses tests continued to rise for 10 days before reversing the trend.\footnote{This time shift is also consistent with the Italian regional data \cite{delayreg}.} In principle, a couple of days should be added to this delay time, due to the period of infectiveness before the development of symptoms \cite{Xi2020}. However, we have kept the value $\tau=10\,d$ considering that: a) a certain percentage of infected people is diagnosed before symptoms occur; b) the results obtained by using nearby values of $\tau$ were pretty similar.

To cope with the heterogeneous nature of the spread of  COVID-19 disease  across the Italian peninsula (see, e.g., the data in Figure~\ref{fig_ISS1} or the web-pages \cite{protciv,ISS,wiki}), we have introduced a multi-region extension of the basic model (\ref{si2r2}). In more details, assuming that we have divided the country into $r$ regions, the multi-region version of (\ref{si2r2}) reads:
\begin{equation}\label{mrsi2r2}
\left\{ \begin{array}{rcl}
\dot x_i(t)     &=& -\frac{\beta_i}{N_i} x_i(t)y_{i1}(t) + \sum_{j=1}^r \rho_{ij} x_j(t),\\[1mm]
\dot y_{i1}(t) &=& \frac{\beta_i}{N_i} x(t)y_{i1}(t)-\sigma_i y_{i1}(t-\tau) s_+(y_{i1}(t))-\gamma_{i1} y_{i1}(t)+ \sum_{j=1}^r \rho_{ij} y_{j1}(t),\\[1mm]
\dot y_{i2}(t) &=& \sigma_i y_{i1}(t-\tau) s_+(y_{i1}(t))-\gamma_{i2} y_{i2}(t),\\[1mm]
\dot z_{i1}(t) &=& \gamma_{i1}y_{i1}(t)+ \sum_{j=1}^r \rho_{ij} z_{j1}(t),\\[1mm]
\dot z_{i2}(t) &=& \gamma_{i2}y_{i2}(t),  \hspace{5cm} \qquad i=1,\dots,r,
\end{array}\right. 
\end{equation}
where $x_i, y_{i1}, y_{i2}, z_{i1}, z_{i2}, \beta_i, \sigma_i, \gamma_{i1}, \gamma_{i2}$ are related to the $i$-th region, and are defined as for (\ref{si2r2}). As before,  we set $\gamma_{i1} = \gamma_1$ as in (\ref{gamma1}).  We see that
$$N_i \, :=\, N_i(t) = x_i(t)+y_{i1}(t)+y_{i2}(t)+z_{i1}(t)+z_{i2}(t), \qquad i=1,\dots,r,$$
is the (time dependent) number of individuals in the $i$th region, whereas the {\em migration coefficients} $\rho_{ij}$ satisfy
\begin{equation}\label{roij}
\rho_{ij}\ge 0, \quad i\ne j, \qquad \sum_{i=1}^r \rho_{ij}=0,
\end{equation}
due to the fact that $\rho_{ij}$, $i\ne j$ is the coefficient of migration from region $j$ to region $i$. We observe that, because of (\ref{roij}), the sum of all right-hand sides in (\ref{mrsi2r2}) vanishes. As a result, one has that
\begin{equation}\label{Nn}
N := \sum_{i=1}^r N_i(t) \equiv const,
\end{equation}
which is the analogue of (\ref{N}) for the multi-region extension of the model.

For our purposes, we have divided Italy into the following $r=4$ macro-regions, depending on the onset of the epidemic and the geography (see Figure~\ref{fig_ISS1}),
\begin{itemize}
\item Lombardy;
\item North, including: Emilia Romagna, Friuli-Venezia Giulia, Liguria, Piemonte, Trentino-Alto Adige, Valle d'Aosta, Veneto;
\item Center, including: Abruzzo, Lazio, Marche, Toscana, Umbria;
\item South (and islands), including:  Basilicata, Calabria, Campania, Molise, Puglia, Sardegna, Sicilia.
\end{itemize}

\section{Results}\label{results} 
 Starting from 24 February 2020 (initial day $t_0=0$), the number of diagnosed active cases $y_{i2}(t_n)$ and the removed = recovered+deceased individuals $z_{i2}(t_n)$ at day $t_n$, published on a daily basis by the Italian Civil Protection Department  \cite{protciv}, have been exploited to tune the parameters of the model (\ref{mrsi2r2}) in order to fit the data in the $i$th macro-region. The fitting procedure, carried out with the aid of the Matlab\,\textsuperscript{\textregistered}  optimization toolbox,\footnote{It is also worth mentioning that the numerical integration of the delay equations (\ref{mrsi2r2}) has been done by using a high-order Gauss-Legendre Runge-Kutta method used as spectral method in time \cite{ABI2020}, which has allowed to greatly reduce the computational cost of the procedure, w.r.t. the use of the Matlab\textsuperscript{\textregistered}  function {\tt dde23}. } has been split into three phases: 
\begin{itemize}
\item[(a)] at first, in each macro-region we fit the corresponding total cases, $y_{i2}(t_n)+z_{i2}(t_n)$, given by the diagnosed active cases and removed people. Adding the third and the last equations in (\ref{mrsi2r2}), we see that the removal coefficient $\gamma_{i2}$ are not explicitly involved at this stage. Moreover, in this fase we set the migration coefficients to 0, i.e., we at first neglect the inter-regions movements; 
\item[(b)] the second optimization phase adapts the solution components $y_{i2}(t_n)$ (positive active cases) and removed  (recovered or deceased) individuals $z_{i2}(t_n)$ in the $i$th macro-region, to the observed data by suitably tuning the removal coefficient $\gamma_{i2}$;
\item[(c)] finally, a further improvement is gained by making the migration coefficients $\rho_{ij}$ in (\ref{mrsi2r2}) come into play.
\end{itemize}
The rationale behind this  approach is to better exploit the degree of regularity of data and make the resulting minimization algorithm more efficient. In fact, in each macro-region,  the total cases exhibit a more regular temporal distribution and thus, in performing phase (a), they play a more important role in tuning all the free parameters except $\gamma_{i2}$.  

The task of splitting the total cases into the two classes of positive active cases, $y_{i2}(t_n)$,  and removed cases, $z_{i2}(t_n)$, is assigned to phase (b). Here, the data display a much more irregular behavior especially in Lombardy, probably due to the human interference in deciding when an infected individual is to be considered completely recovered. As a matter of fact,  a huge amount of positive cases moving from active to recovered may be observed in some specific days. This is mainly due to the criteria of discharge of COVID-19 patients adopted in Italy and, in particular, to the ``negativization tests'' for home quarantined people  (hospitalized discharges are less affected by this phenomenon). 
For this reason, the  coefficients $\gamma_{i2}$ have been assumed piecewise-constant on time intervals of length of (at least) $30\,d$.  

Lastly, phase (c) refines the fitting of the solution to the data: it has played a significant role during the days approaching the lockdown which saw a considerable number of workers and students, native to South Italy, to leave the northern regions in order to avoid the quarantine restrictions imposed by the Italian government.

\begin{figure}[t]
\centerline{\includegraphics[width=5.75cm]{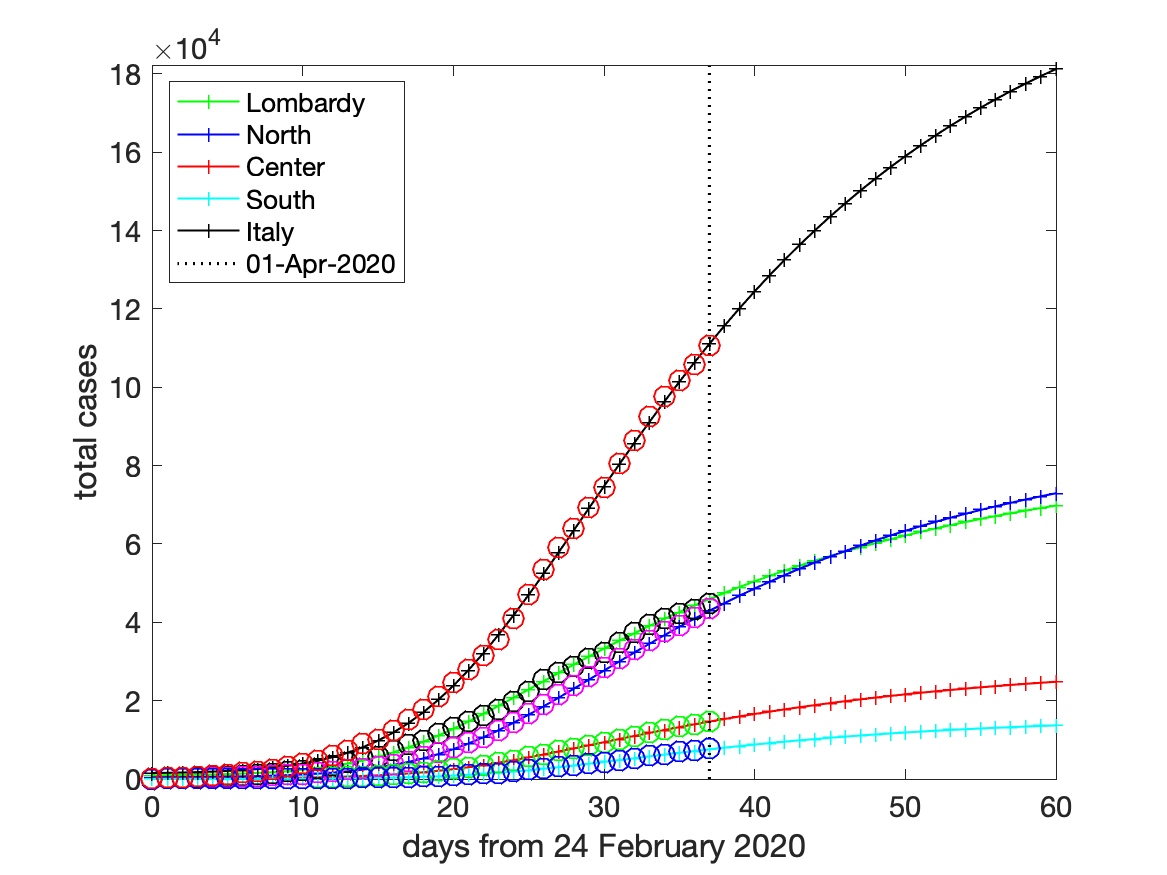}\quad \includegraphics[width=5.75cm]{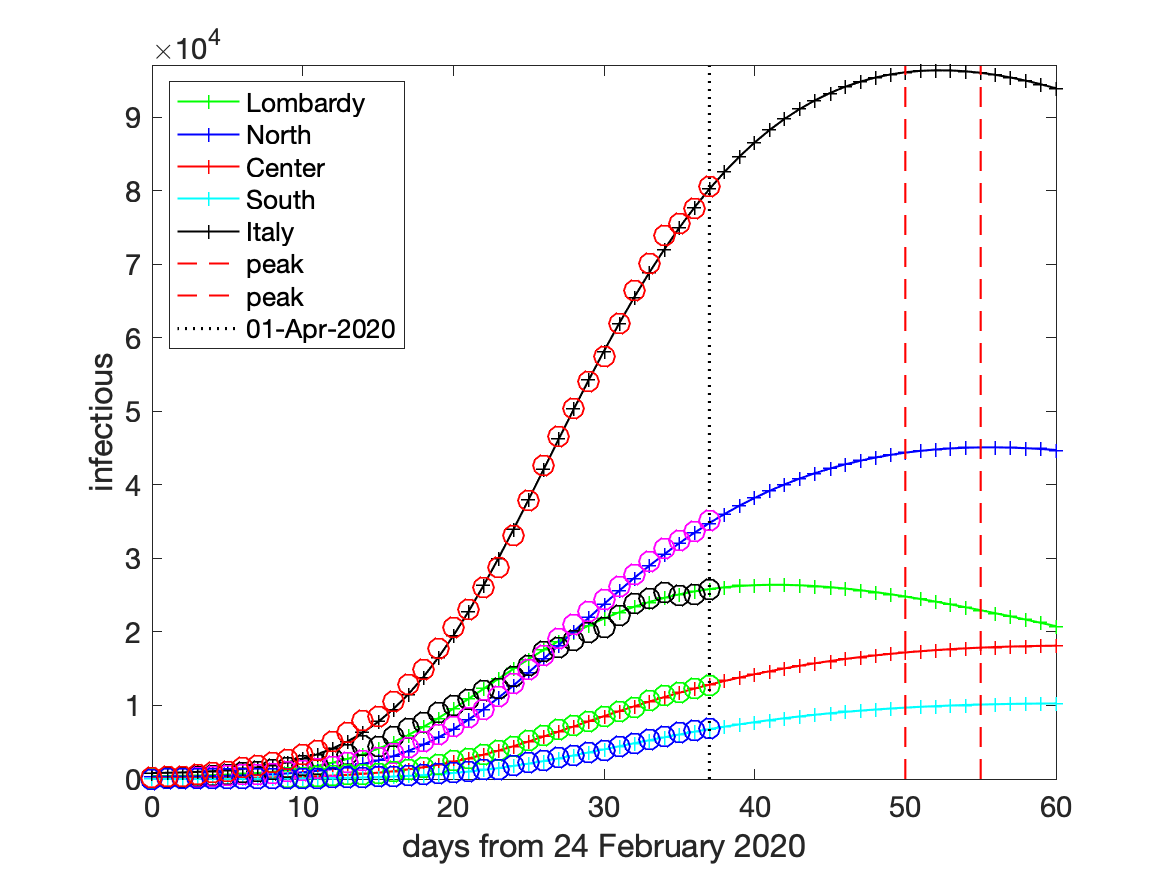}}

\medskip
\centerline{\includegraphics[width=5.75cm]{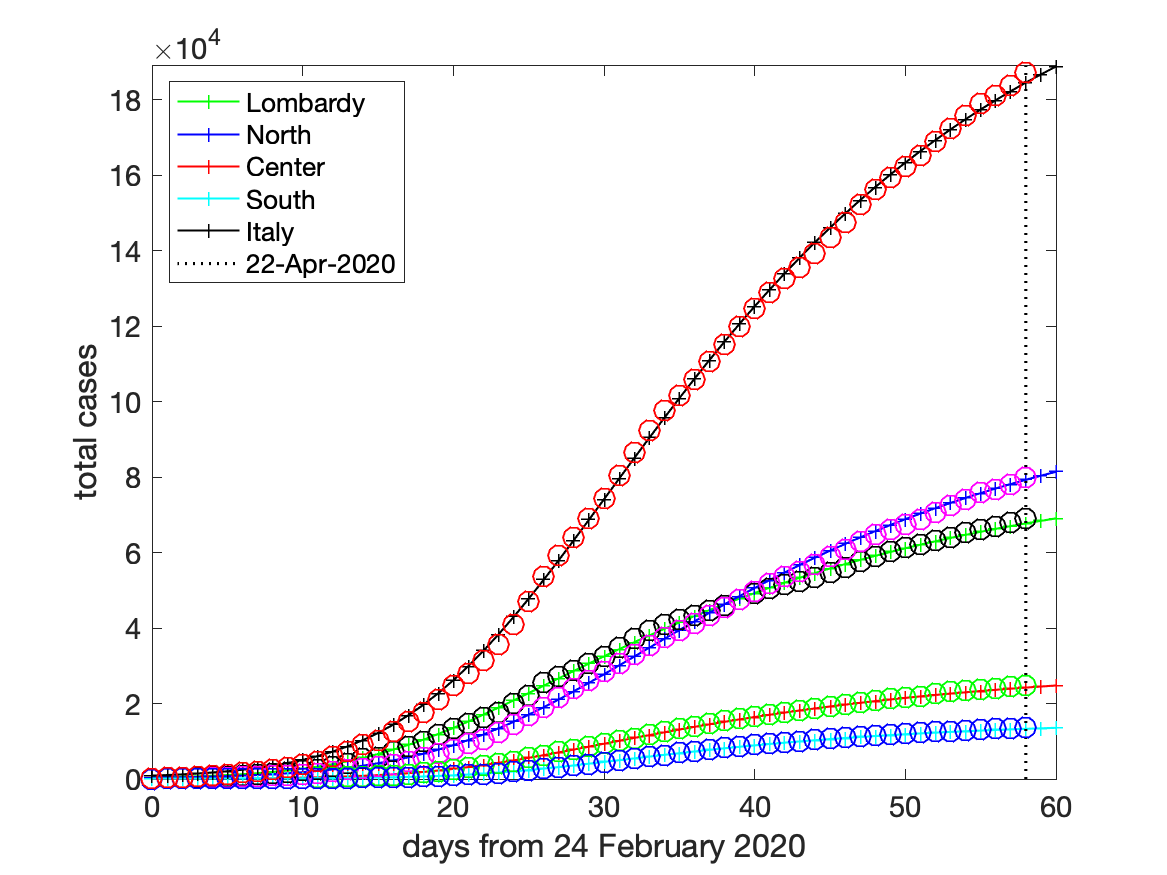}\quad \includegraphics[width=5.75cm]{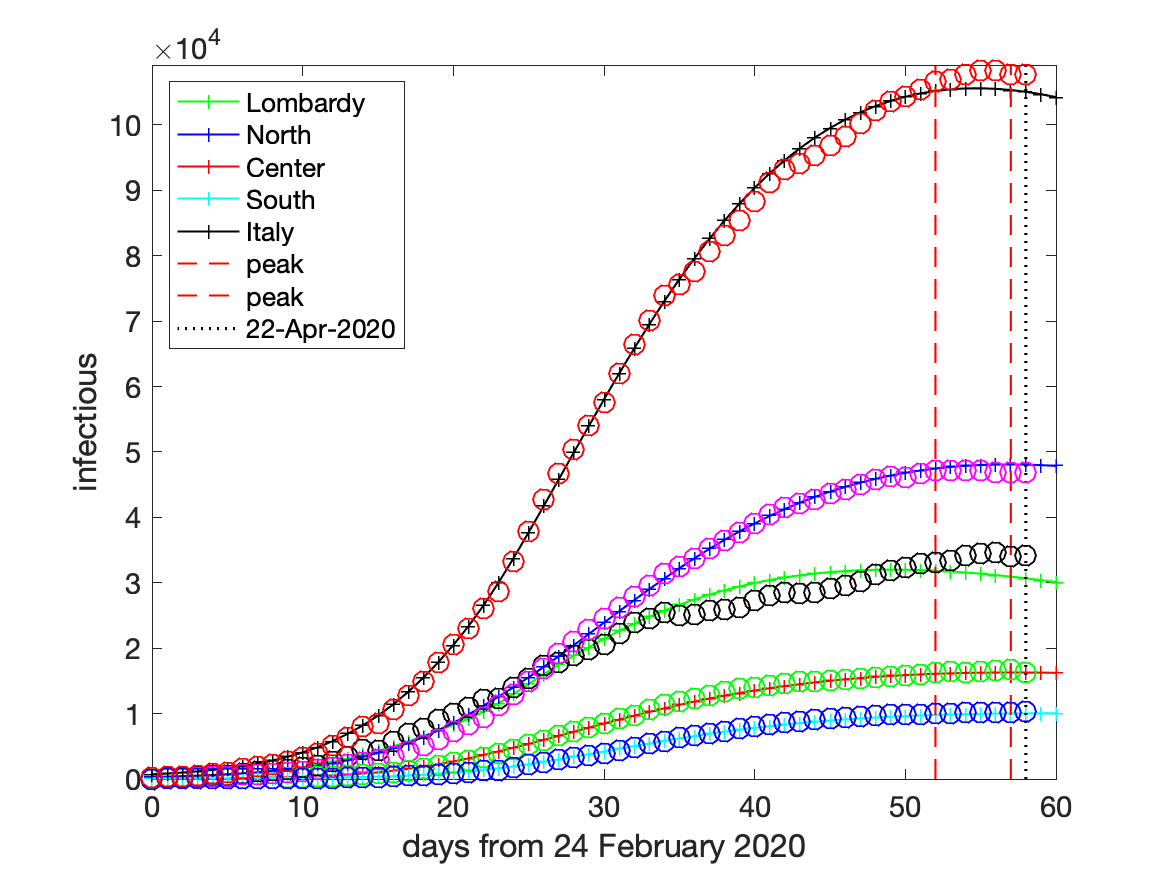}}
\caption{forecast for the total cases  (left pictures) and diagnosed active cases (right pictures) on  1 and 22 April 2020, respectively.}
\label{fig1}
\end{figure}

The results for the whole country are obtained by summing those in each macro-region. To this end,  hereafter we set:
\begin{equation}\label{Italia}
x(t) = \sum_{i=1}^4 x_i(t), \qquad y_j(t) = \sum_{i=1}^4 y_{ij}(t),\quad  z_j(t) = \sum_{i=1}^4 z_{ij}(t),\quad j=1,2,
\end{equation}
representing the 5 classes of the model related to Italy.

\subsection{Reliability of the model}

Forecasts of the total and active confirmed positive cases, in the above four macro-regions and Italy, are generated every day and collected at the web-site \cite{mrsir} starting from 1 April 2020. Here we report a few of them, in order to show the reliability of the model. 

\paragraph{Detection of the pick of confirmed active cases.}
On April 19, the curve of the Italian active (diagnosed) cases $y_{2}(t_n)$ attained its maximum of 108257 individuals, and exhibited a flat slope in the neighbouring days before slowing down. To check the accuracy of the model in predicting the peak day and the number of total confirmed cases $y_{2}(t_n)+z_{2}(t_n)$, we have solved the model by fitting its solutions to the observed data available on April 1 and 22. The results for these two datasets are displayed in the top and bottom pairs of pictures in  Figure~\ref{fig1}, respectively.   
The solid lines are predicted by the model, whereas the circles correspond to the observed data. The dashed vertical lines locate the days around the predicted peak where the number of active cases differ at most of $0.4\%$ from the maximum. We see that both the total cases and the peak day in Italy (corresponding to day 55) are quite well-predicted  on a time range of more than twenty days.

\paragraph{Estimation of the actual attack rate.}
It is widely believed that the number of infected people who remain hidden to  COVID-19 testing programs is likely far grater than the number of confirmed cases worldwide. The importance of practicing social distancing and safety measures not only reflects the risks associated with the pre-symptomatic stage of the disease, but also suggests that an important role in the transmission of the virus might be played by asymptomatic and pauci-symptomatic people.  In absence of a population-wide testing, it is difficult to estimate how many people get infected without showing significant symptoms. As for now, there are isolated studies which cannot be extended to general contexts, even though a percentage of asymptomatic carriers  reaching the  80\% among all infected people has been considered a possible outcome in areas with high prevalence of circulating infection \cite{InCoGr2020}.  

\begin{figure}[t]
\centerline{\includegraphics[width=5.75cm]{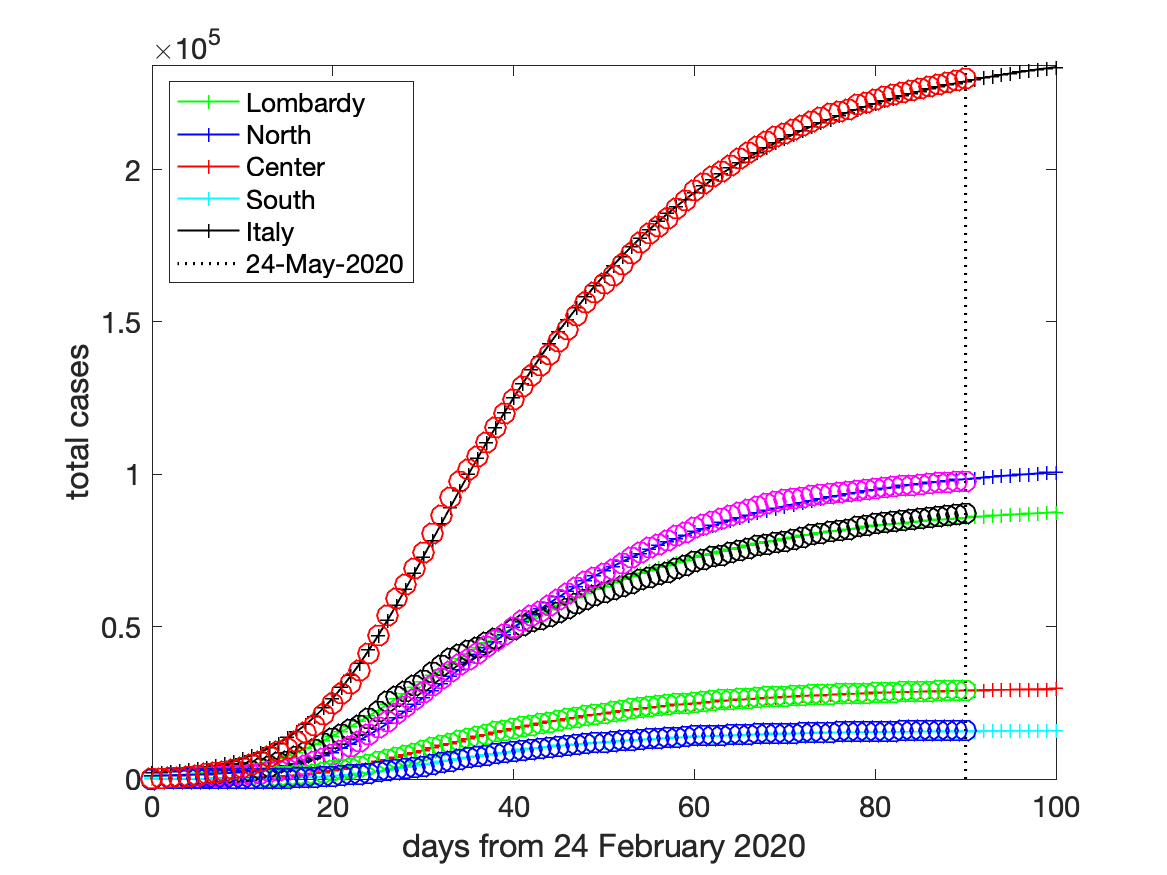}\quad \includegraphics[width=5.75cm]{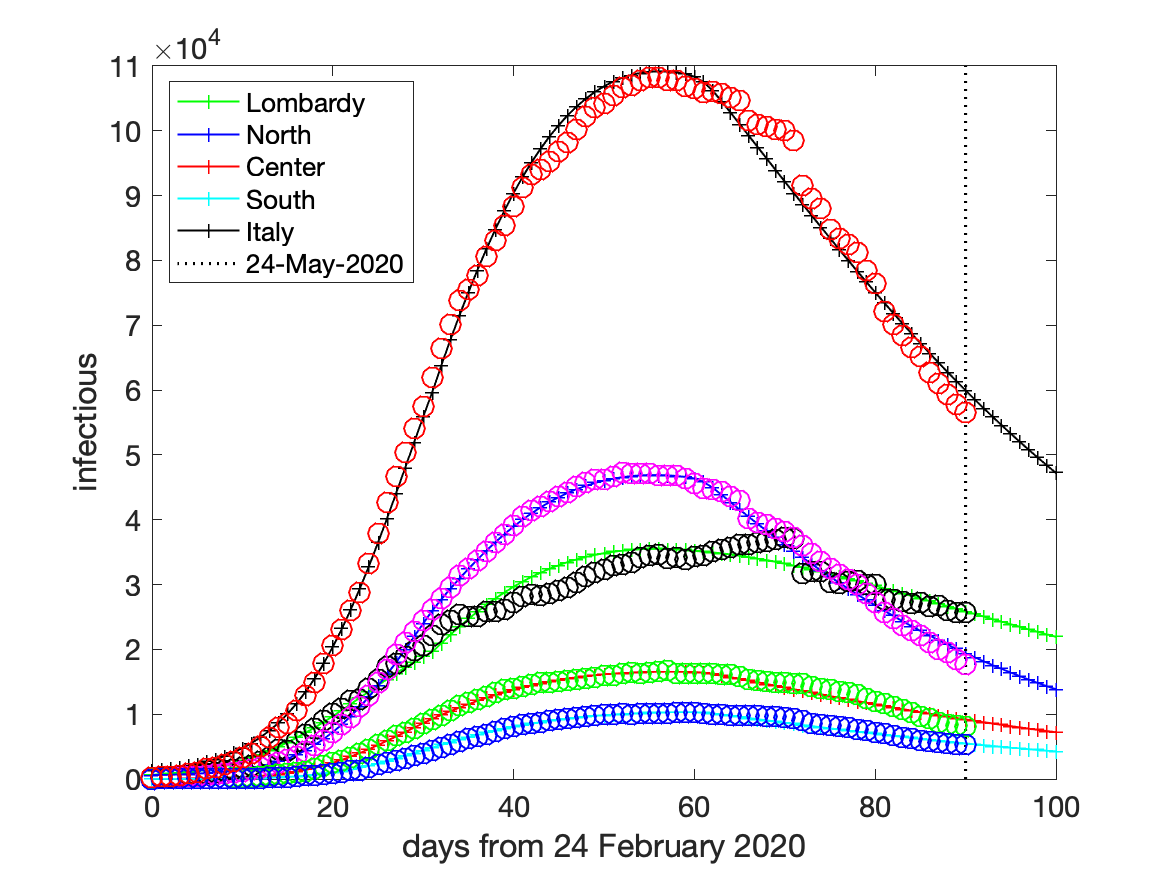}}

\medskip
\centerline{\includegraphics[width=5.75cm]{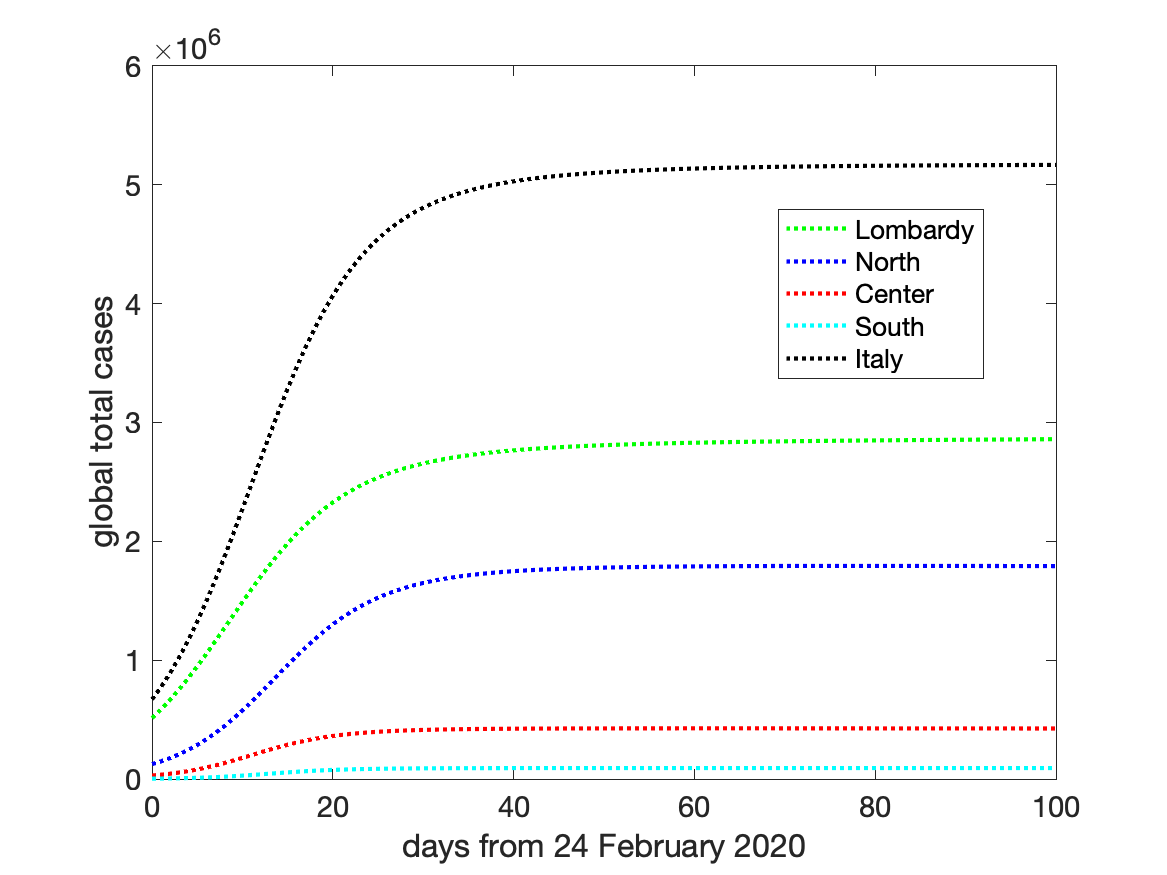}\quad \includegraphics[width=5.75cm]{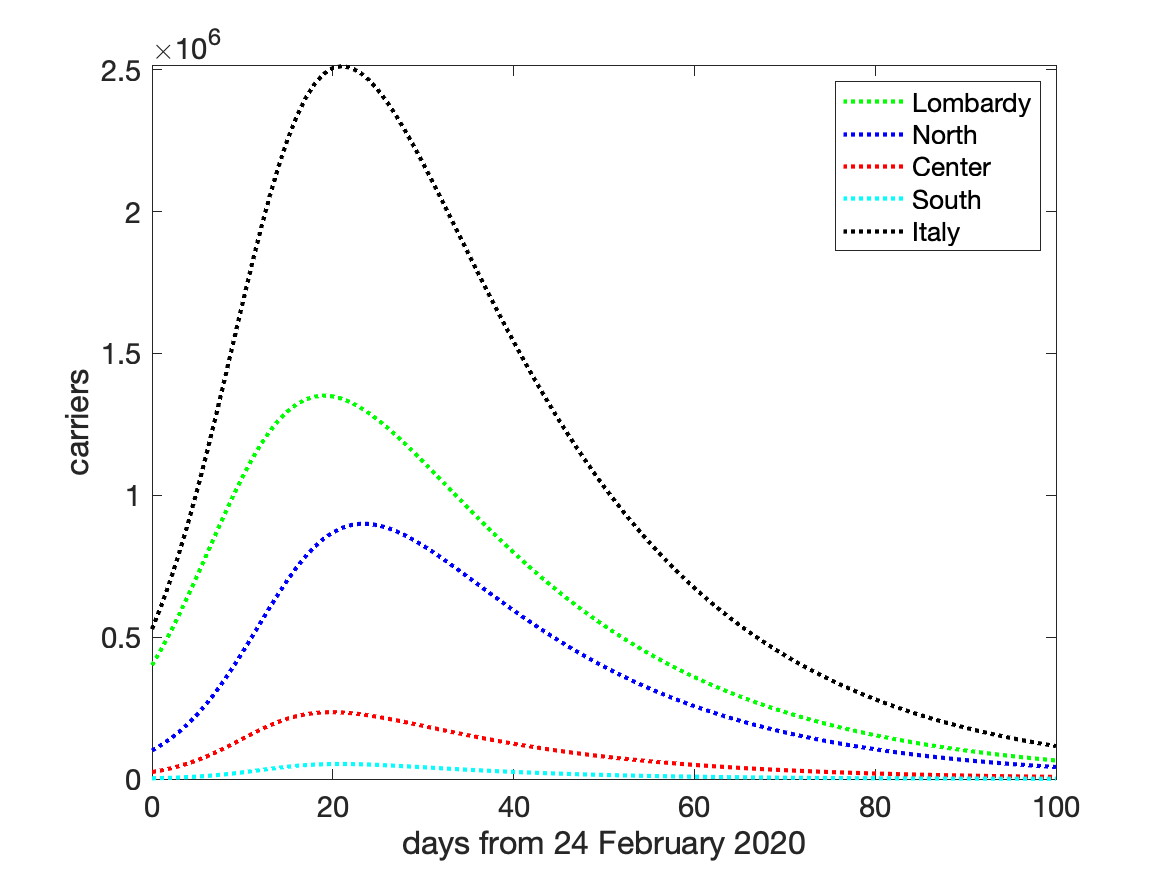}}
\caption{Top: simulated and actual data on 24 May 2020 (day 90) for the total cases (left plot) and diagnosed active cases (right plot). Bottom: estimated global total cases and carriers using the parameters of 24 May 2020 (day 90).}
\label{fig_d90}
\end{figure}

While waiting for an extensive screening program, one route for investigating the number of hidden infections is to rely on  a suitable compartmental epidemiological model, able to explain the dynamics of  virus shedding in observed data and to yield possible scenarios of those classes of population that may not be directly traced  (see, for example, \cite{FaPi2020,XiFu2020,Gior2020}). By considering the class $y_{1}(t)$ of untested infected people, our model offers a tool to to get an estimation of the portion of population who actually got the virus (actual attack rate). We emphasize that, strictly speaking, our model does not consider a separate class for asymptomatic people, but we can get some insight into this additional question under suitable assumptions which depend on the context (see below for a discussion).

As a reference dataset to tune the parameters of the model, we used the observed data updated to 24 May 2020 (day 90).\footnote{This dataset will be also considered for later use.} The aim of the two upper pictures in Figure~\ref{fig_d90} is to assess the accuracy of the model in fitting the data. In particular:
\begin{itemize}
\item the upper-left picture  shows the total confirmed cases (circles) and the corresponding solutions $y_{i2}(t_n)+z_{i2}(t_n)$, $i=1,\dots,4$, for the four macro-regions and $y_{2}(t_n)+z_{2}(t_n)$ for the whole Italy obtained in output by the model (solid lines). We see that the yielded accuracy is quite good, and is fostered by the regular course, from the appearance of the symptoms to the disease detection, ruling the transition from the class $y_{i1}$ of infected untested people to the class $y_{i2}$ of confirmed patients;

\item the upper-right picture shows the active confirmed  cases (circles) and the corresponding solutions $y_{i2}(t_n)$, $i=1,\dots,4$, for the four macro-regions and $y_{2}(t_n)$ for the whole Italy. In this case, the data display a  certain irregularity especially in Lombardy and in the rest of Northern Italy. As was discussed above in the description of the optimization algorithm, here the transition from illness (class $y_{i2}$) to healing (class $z_{i2}$) heavily depends on human decision-making. In any event, the regular shape of the model solution curves  describe the overall behavior of the data with good enough accuracy.       
\end{itemize} 

\begin{table}[t]
\caption{diagnosed and global total cases after the end of the lockdown.} 
\label{tab1}
\centerline{\small \begin{tabular}{|c|r|r|r|c|c|}
\hline
                 &\multicolumn{2}{c|}{diagnosed}  &\multicolumn{1}{c|}{global} &global/ & immunity\\
                 & measured & simulated &simulated  & diagnosed ratio & rate\\
\hline
Lombardy & 78105  & 78910 &  2841561 &     36  &      28.2\%    \\
North        & 91600  & 89544  &  1792530 &    20 &   10.1\%    \\
Center       & 27205  & 26948  &  426153 &    16 &    3.20\%    \\
South        & 15028   & 14853  &  92027  &      ~\,6 & 0.48\%  \\
\hline
Italy           & 211938 &210256 &  5152270 & 25&    8.53\%     \\
\hline
\end{tabular}}

\bigskip
\caption{peak of the carriers.} 
\label{tab2}
\centerline{\begin{tabular}{|c|cccc|c|}
\hline
       & Lombardy & North & Center & South & Italy \\
       \hline 
day   & 19 & 23 & 20 & 21 & 21 \\
level & $1.4\cdot 10^6$  &  $9.0\cdot 10^5$ &   $2.4\cdot 10^5$  &  $5.3\cdot 10^4$ &  $2.5\cdot 10^6$\\ 
\hline
\end{tabular}}
\end{table}

After selecting the parameters, we can look at the other components of the solution which reveal the {\em hidden side} of virus diffusion in Italy. They are displayed in the two bottom pictures of Figure~\ref{fig_d90}:
\begin{itemize}
\item the bottom-left picture shows the number of the overall infected individuals, namely the total cases including both diagnosed and untested people. These are easily computed  as the sums $y_{i1}(t)+z_{i1}(t)+y_{i2}(t)+z_{i2}(t)$ 
in each macro-region, and their sum for the whole Italy. The model reveals a very dramatic scenario with an actual attack rate  far greater than the one corresponding to the confirmed cases. Table~\ref{tab1} summarizes the number of these global cases right after the end of lockdown on May 3, 2020, and compare them with the confirmed ones; 

\item the bottom-right picture shows the evolution in time of carriers, namely the undiagnosed active cases forming the classes $y_{i1}(t)$ for each macro-region and, according to (\ref{Italia}), $y_1(t)$ for Italy. They exhibit a behavior quite different from that of the diagnosed active cases (upper-right plot in Figure~\ref{fig_d90}). The most evident difference concerns  the peak day which occurs about one month earlier than the corresponding one in the diagnosed active cases.  As is expected, it falls in proximity of the nation lockdown on March 11 (day 16).  Interestingly, our simulation reveals how the forced distancing-measures, imposed by the government to all the nation, impacted on the infection growth curve which, after a few days, reached its maximal level and rapidly decreased during the subsequent days. Table~\ref{tab2} reports the day and the level of the peak in each macro-region and in Italy.     
\end{itemize}
 
The extent to which these percentages may reflect the real epidemic diffusion in Italy will be revealed once a screening of a wide random sample of  population will be carried out. In any event, Table~\ref{tab1} discloses very different scenarios for the four analyzed macro-areas, depending on the onset of the infection.

\begin{table}[t]
\caption{increase of mortality from 20 February to 31 March 2020.} 
\label{tab3}
\centerline{\small \begin{tabular}{|c|c|c|c|c|}
\hline
&                                   & average                      &deaths                 &  additional             \\
&  total deaths              & total deaths                 &COVID-19           & increase of\\
& \multicolumn{1}{c|}{2020} & \multicolumn{1}{c|}{2015-19} &\multicolumn{1}{c|}{2020} & deaths 2020     \\
\hline
Lombardy     & 27279  & 11195  &  8362    &  7722      \\
North        & 29123  & 21296  &  4195    &  3632      \\
Center       & 13985  & 13120  &  749     &  116       \\
South        & 20559  & 19981  &  404     &  174       \\
\hline
Italy           & 90946 &65592 &  13710 & 11644\\
\hline
\end{tabular}}
\end{table}

In the north of Italy, where the infection growth rate reached dramatic levels, the local healthcare system became overwhelmed. Consequently, a relevant number of infected people, even those experiencing symptoms, were supposed to escape any test detection. In support of this commonly shared hypothesis, one may observe that, from late February through all of March, there was a huge amount of not-counted deaths  which may indirectly be related to SARS-Cov2. In Table\,\ref{tab3}, we summarize some results coming from a study performed by ISTAT (Italian Institute of Statistics) and ISS, dated 4th of May, which compares the total deaths from 20 February to 31 March 2020 (excluding 29 February 2020)   with the yearly average deaths of the past 5 years.  From the last column in Table\,\ref{tab3}, we see that a huge excess of morality, not ascribed to the virus lethality, has been experienced in this period by the northern regions. Among these are certainly deaths due by poor health conditions that might normally have been treated, if hospitals had not collapsed by a surge of patients needing intensive care.  However, considering that deaths for and with COVID-19 were counted only for those people who tested positive before passing away,  it is not unfounded to argue that a considerable amount of such a discrepancy might be related to virus infection which, in turn, suggests a certain level of correlation between this excess of deaths and the actual proportion of population who has been infected without being diagnosed. 
In this respect, Table~\ref{tab1}, shows that the estimated cumulative number of infected  are about 25 times the confirmed ones in the whole Italy, and even more  in Lombardy,  where an immunity rate of about 30\% seems to be achieved at the end of lockdown.

On the contrary, in the Center and South of Italy the virus had a minor impact and the public emergency measures were introduced in time, so that the spread was successfully contained. In particular, in the South of Italy, where 
the healthcare system  did not go out of control,  we can speculate that a great deal of untested infected people might actually be asymptomatic or pauci-symptomatic. Under this simplifying assumption, from Table~\ref{tab1} we can guess that more than 80\% of the simulated $92027$ individuals in the South of Italy could be mildly symptomatic. This percentage can make sense, also considering that a certain number of asymptomatic individuals is anyway detected by means of contact-tracing policies, thus entering the class $y_{42}$ of confirmed cases.

Finally, we stress that the results illustrated above apply for the specific choice of the two static parameters $\gamma_1$ and $\tau$ in the model. A sensitivity analysis shows that, while increasing or decreasing  $\tau$ of a couple of days does not produce considerable changes in the results, a variation of  $\gamma_1$ seems to affect the solution more directly. In particular, diminishing the value  of this parameter would result in an increase of the estimated values of the total cases so that, in this event, our results would underestimate the actual situation.


\subsection{The lockdown effect}

\begin{figure}[ht]
\centerline{\includegraphics[width=11.8cm]{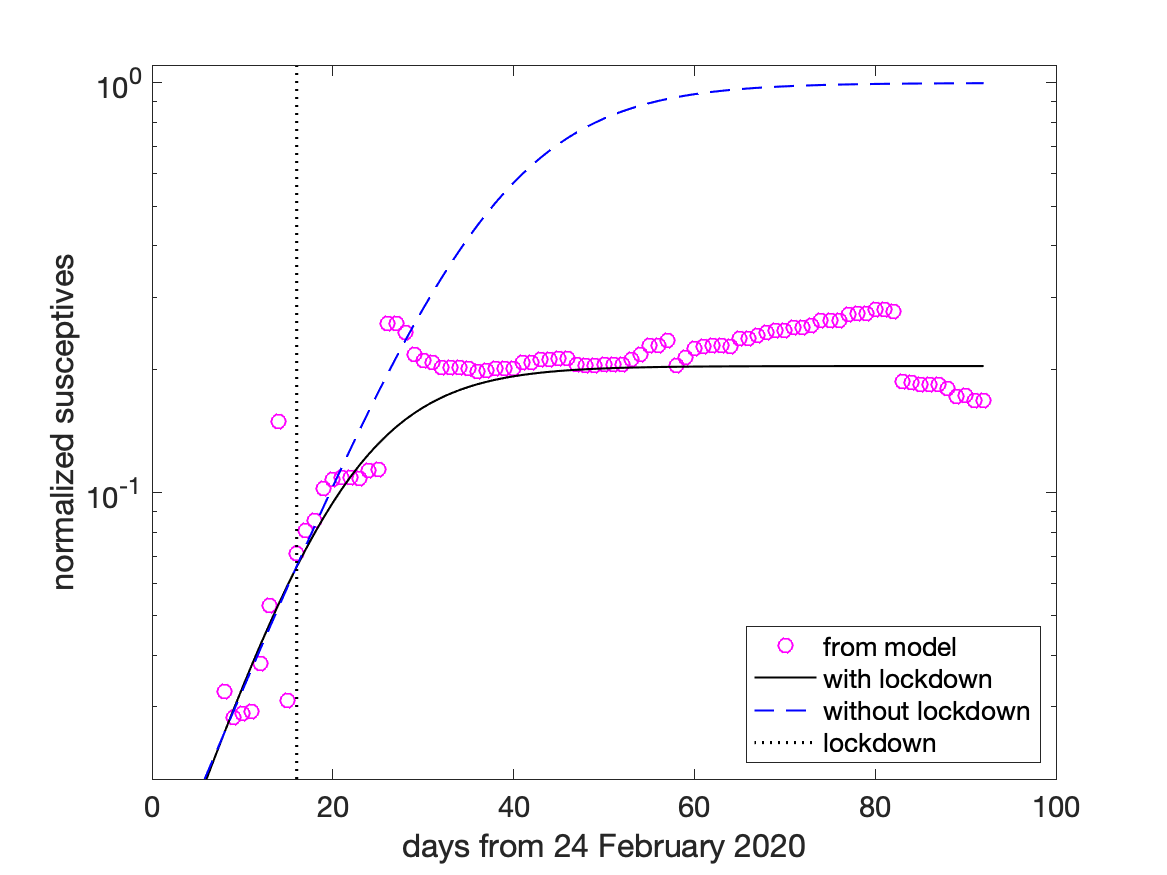}}
\caption{scenario with and without lockdown in North Italy (including Lombardy).}
\label{fig2}
\end{figure}

An interesting use of the model is to ascertain the benefits of lockdown, which was imposed by the authorities starting from 11 March 2020 (day 16 starting from February 24). At that time, the spread of the disease in the regions of Center and South Italy was contained, and the data were scarce. Consequently, in our analysis of the pre-lockdown period (actually, before day 24), we have discarded the data from regions in the Center and South plus islands. 

As was outlined above, the model (\ref{mrsi2r2}) derives an estimate for the initial values of the susceptives,  $x_i(0)$, $i=1,\dots,4$ (and thus for $x(0)$, according to (\ref{Italia})), during the fitting phase (a). As one would expect, this number increases during the initial part of the spread of an epidemic, reflecting the space diffusion of the disease in new uninfected populated areas, until it eventually saturates the whole population of a given region.  

In Figure~\ref{fig2} the circles are the number of susceptives in North Italy (including Lombardy) estimated by the model (\ref{mrsi2r2}), normalized by the whole population (about $2.7\cdot 10^7$ people), with the solid line being its logistic approximation. Because of the lockdown, which is marked by the vertical dotted line at day 16, one infers that about 20\% of the whole population becomes eventually exposed to the virus. We see that, after the lockdown, the estimated susceptives continue to increase for about 10-12 days, which is consistent with the maximum period of incubation of the virus. After that, the number of susceptives stabilizes.

To figure out how the scenario would have played out in absence of lockdown, we use the pre-lockdown data to define a new logistic approximation which coincides with the previous one until day 16 (i.e., the day of lockdown), and eventually saturate the whole population. It is the dashed curve in Figure~\ref{fig2}, showing that almost all population becomes susceptible after about 90 days (24 May 2020).  This reasoning allows us to virtually compare the real situation at day 90, with a hypothetical one where all the parameters of the model remain fixed, but the susceptives, which are set equal to the whole population.
The results are summarized in Figures~\ref{fig_d90}, \ref{fig_LNL}, and \ref{fig_LNL1}: 

\begin{itemize}

\item in Figure~\ref{fig_d90}, we displayed the simulated curves (solid lines) and observed data (circles) for the total diagnosed cases and the active diagnosed cases on 24 May 2020 (i.e., day 90). As one may see,  the free parameters may be finely tuned in order to force the solution of (\ref{mrsi2r2}) to lie very close to the observed data in each macro-region and, therefore, in the whole Italy. Excluding the initial {values $x_i(0)$ of susceptible individuals in each macro-region}, the set of parameters obtained by this simulation has then been used to infer the scenario in absence of lockdown; 

\begin{figure}[t]
\centerline{\includegraphics[width=5.75cm]{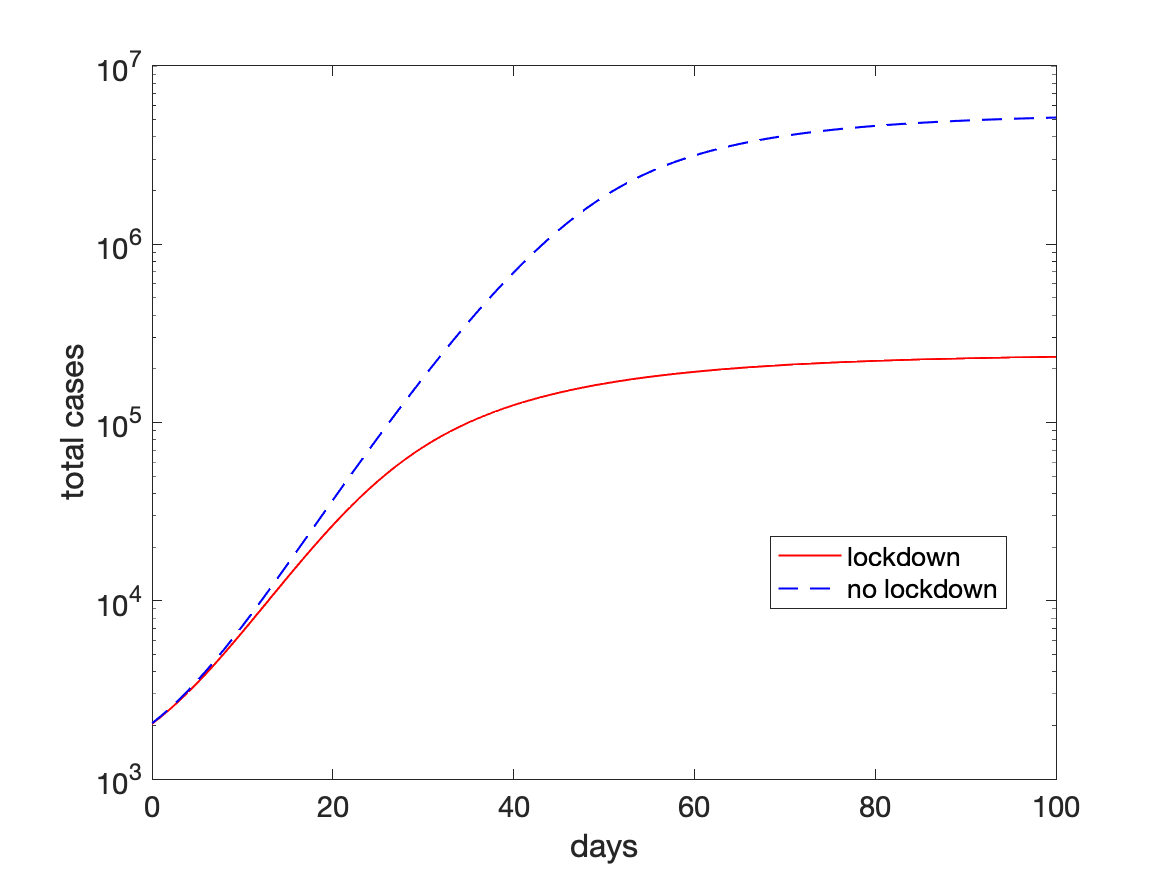}\quad \includegraphics[width=5.75cm]{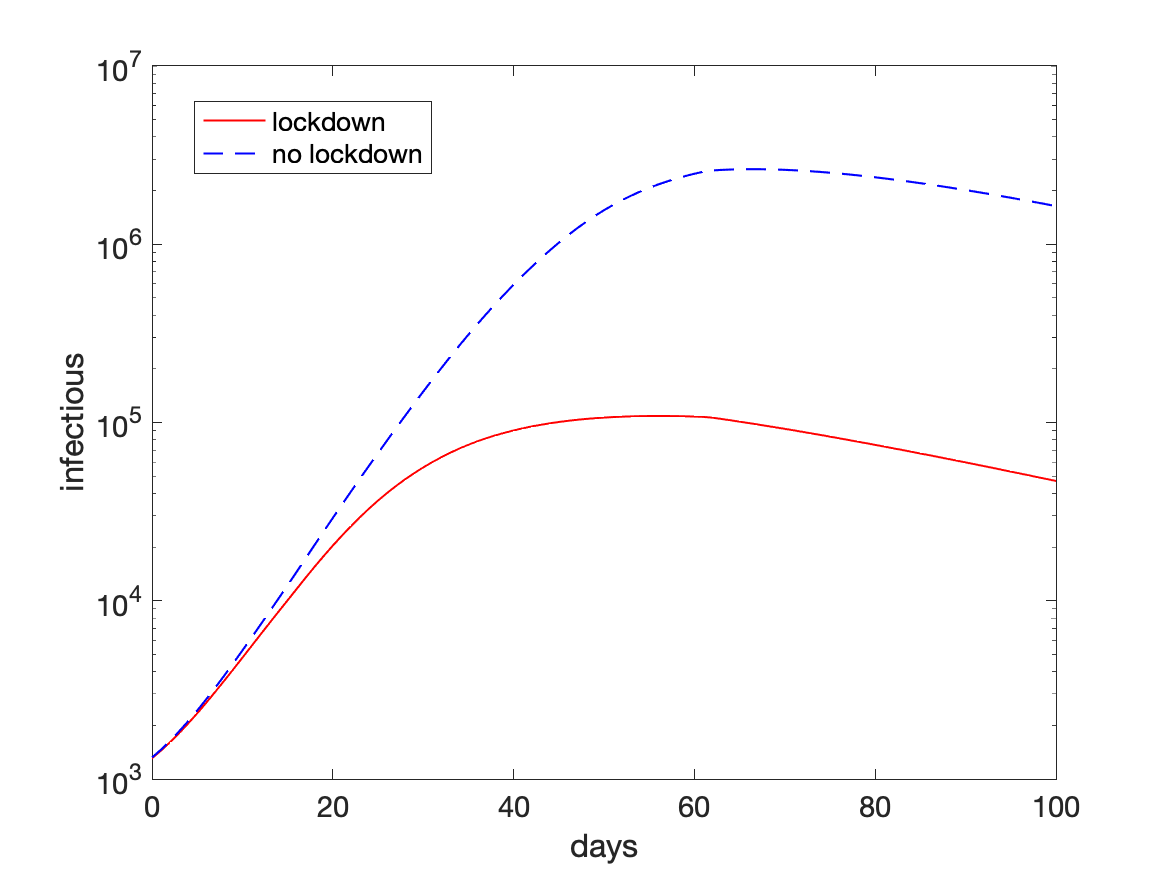}}
\caption{simulated total cases (left plot) and diagnosed active cases (right plot) for the whole Italy with and without lockdown using the parameters of 24 May 2020 (day 90).}
\label{fig_LNL}
\end{figure}

\item in Figure~\ref{fig_LNL}, we report again the total cases and active diagnosed cases for the whole Italy (solid line) in semilogarithmic scale, corresponding to the situation with lockdown, and compare them with the simulation obtained by using all the parameters as in the previous case, but the susceptives, which are set to the whole population in each macro-region (dashed lines). As one may see, at day 100 the forecast for the total cases passes from less than $2.5\cdot 10^5$, with the lockdown, to about $5\cdot 10^6$ in the no-lockdown scenario. Correspondingly, the infectious individuals have a peak of about $10^5$ cases, with lockdown, whereas it is more than $2\cdot10^6$ in the no-lockdown case. Assuming the same lethality for the disease in the two scenarios, this would mean passing from about $3.4\cdot 10^4$ deaths to about $7\cdot 10^5$. It should be emphasized that the no-lockdown scenario assumes that no new kind of emergency measure is introduced during the simulated period by the government or by the population itself, and does not take into account a number of critical variables that would likely affect the obtained results such as, for example, the effects of overloading healthcare structures; 

\begin{figure}[t]
\centerline{\includegraphics[width=5.75cm]{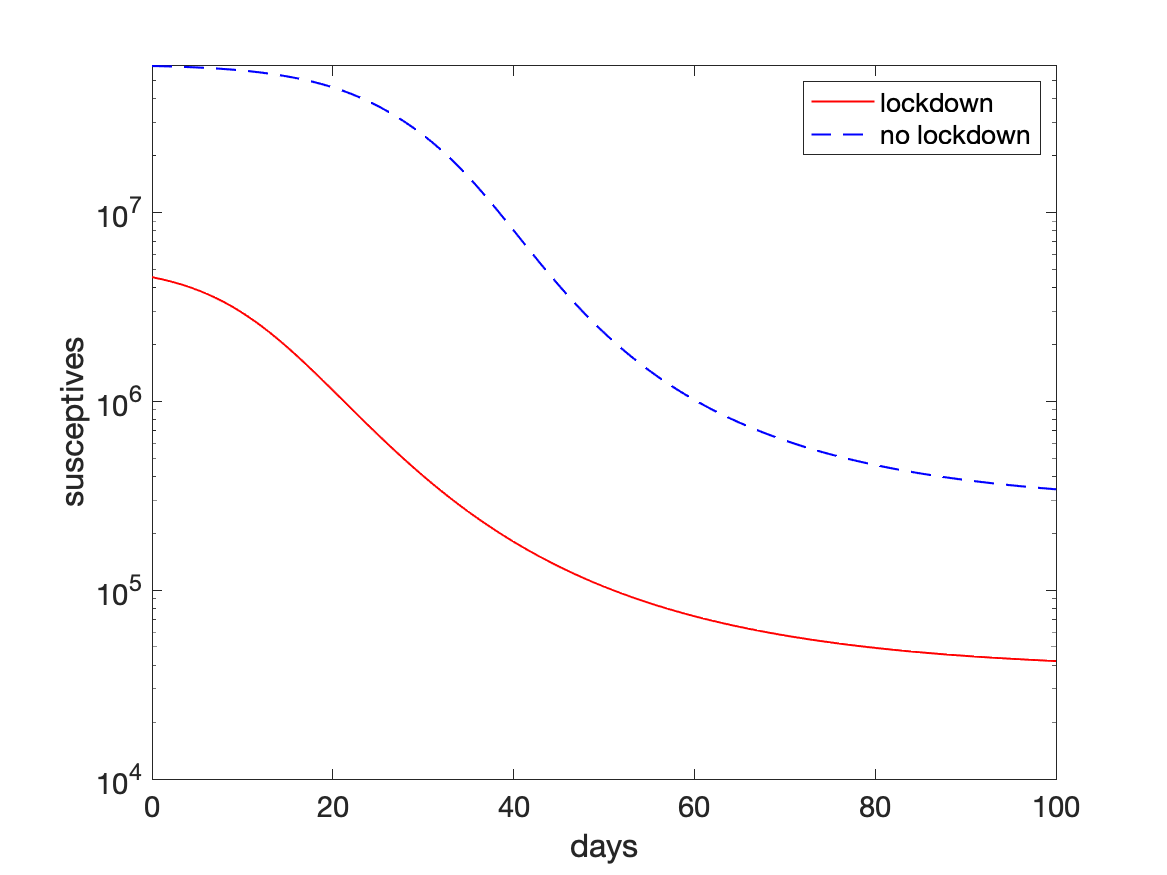}\quad \includegraphics[width=5.75cm]{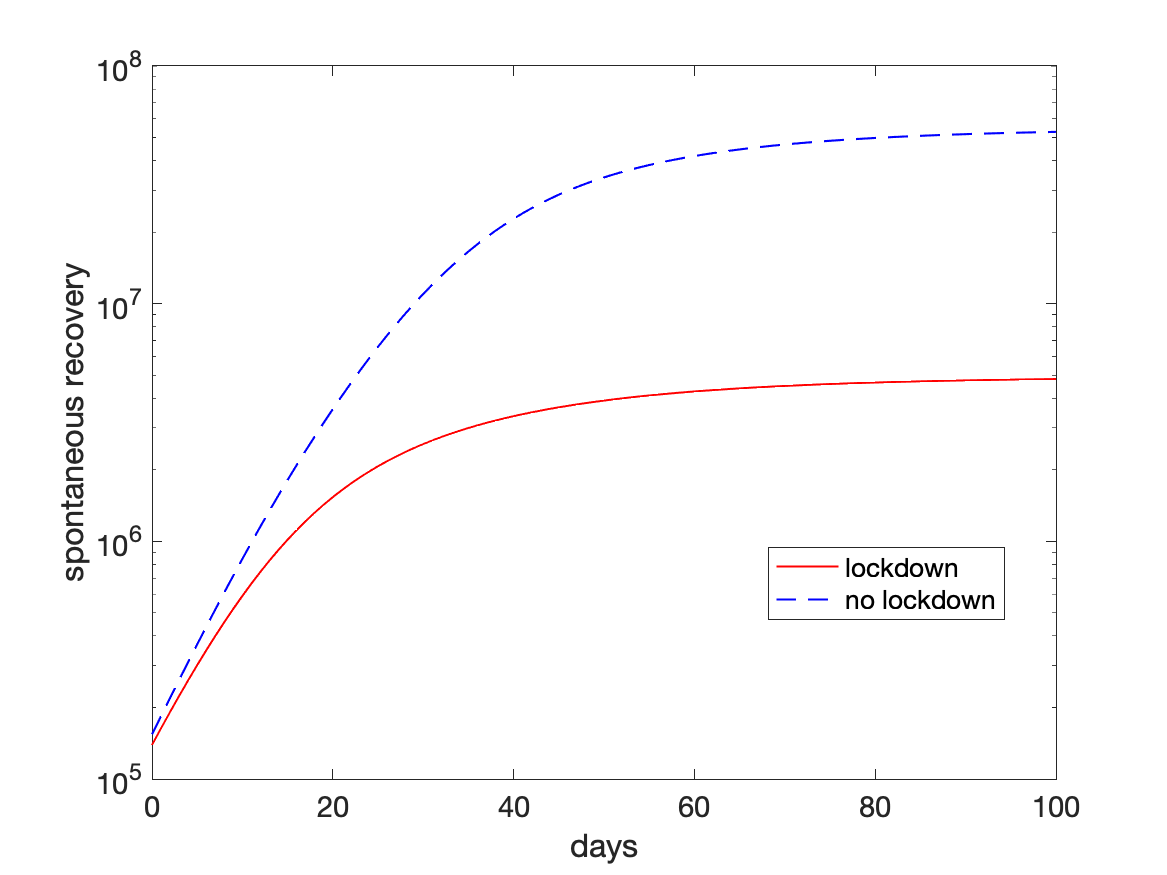}}
\caption{simulated susceptives (left plot) and undiagnosed recovered (right plot) for the whole Italy with and without lockdown using the parameters of 24 May 2020 (day 90).}
\label{fig_LNL1}
\end{figure}

\item in Figure~\ref{fig_LNL1}, it is depicted the simulation of the total susceptives and of undiagnosed infected people spontaneously recovered, i.e. (see (\ref{Italia})), the classes $x(t)$ and  $z_1(t)$, with and without lockdown. As one may infer from the left plot, in the no-lockdown case, one reaches a so-called {\em herd immunity} \cite{RaBa2020}, with most of the population infected by the virus, even though many spontaneously recover, as is shown in the right plot of the figure.

\end{itemize}

\subsection{Anticipating the lockdown}
\begin{figure}[ht]
\includegraphics[width=11.8cm]{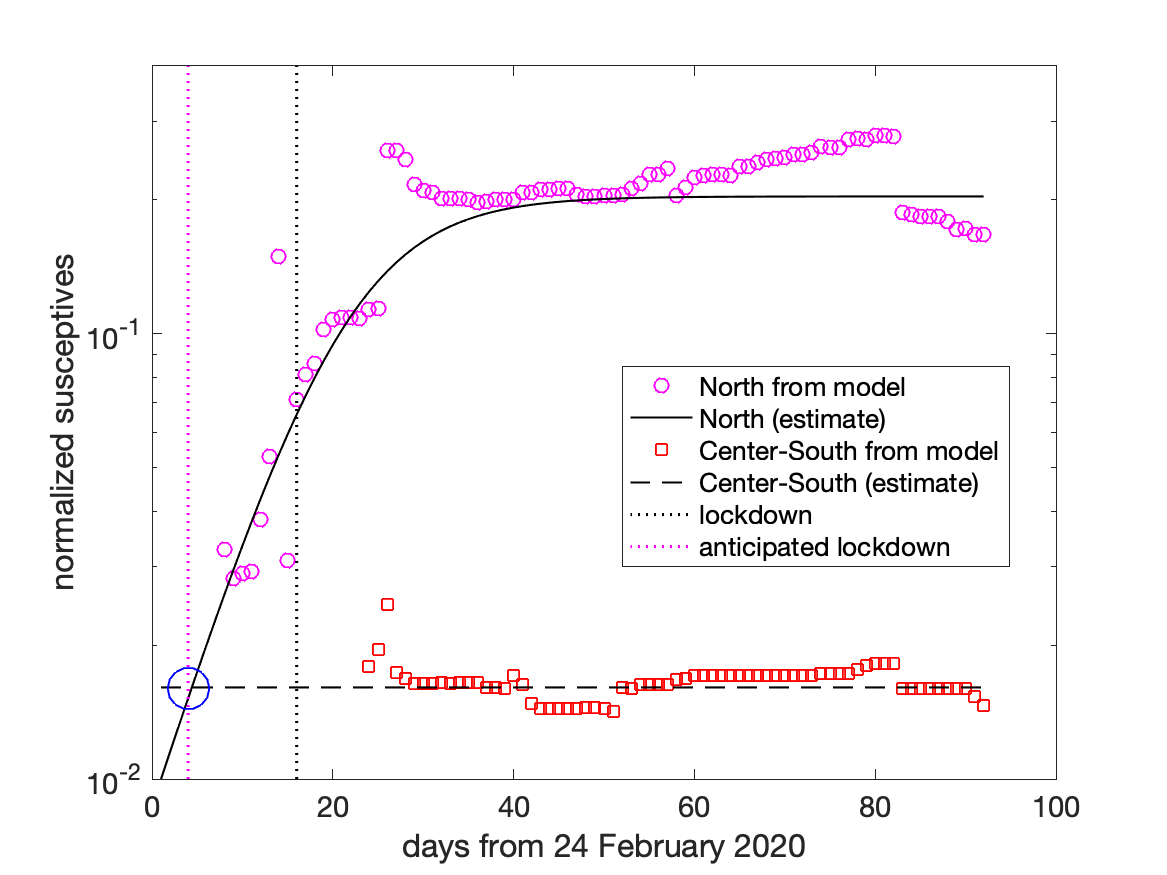}
\caption{Effect of an anticipated lockdown at day 4, instead of day 16.}
\label{fig3}
\end{figure}

As was outlined above, the lockdown was able to stop the epidemic spread in the regions of the Center and South Italy where the contagious arrived later, whereas in the North of Italy the situation has been much more critical, as is testified by the huge number of deaths with respect to the rest of the country.  The question then arises as to when the North Italy should have been locked down in order to share a development of the disease similar to that observed in the other regions.

To this end, in Figure~\ref{fig3} we plot the normalized susceptives estimated by the model in the North Italy (including Lombardy, as we did in Figure~\ref{fig2}), with the circles starting from day 8, and in the Center and South Italy summed together, given by the squares starting at day 24, since in the preceding days the cases where too few to have a sufficient statistical accuracy. Also in this case, the number of the susceptives is normalized by the population in Center and South Italy (about $3.3\cdot 10^7$ people). 

As one may see, now the level of the susceptives exposed to the virus is much smaller (less than 2\%) and this has been the main reason for the better situation in the Center and South of Italy. Consequently, one infers that a similar outcome would have emerged in North Italy, too, provided that the lockdown had started when the level of susceptives was similar. Form the logistic approximation obtained before (solid line in Figure \ref{fig3}), one deduces that this would have been the case with the lockdown starting at day 4, instead of day 16 (see the circle at the intersection of the two estimates). This means that imposing a lockdown in North of Italy two weeks in advance would probably have  guaranteed a much less dangerous spread of the epidemic.   

\begin{figure}[t]
\centerline{\includegraphics[width=11.8cm]{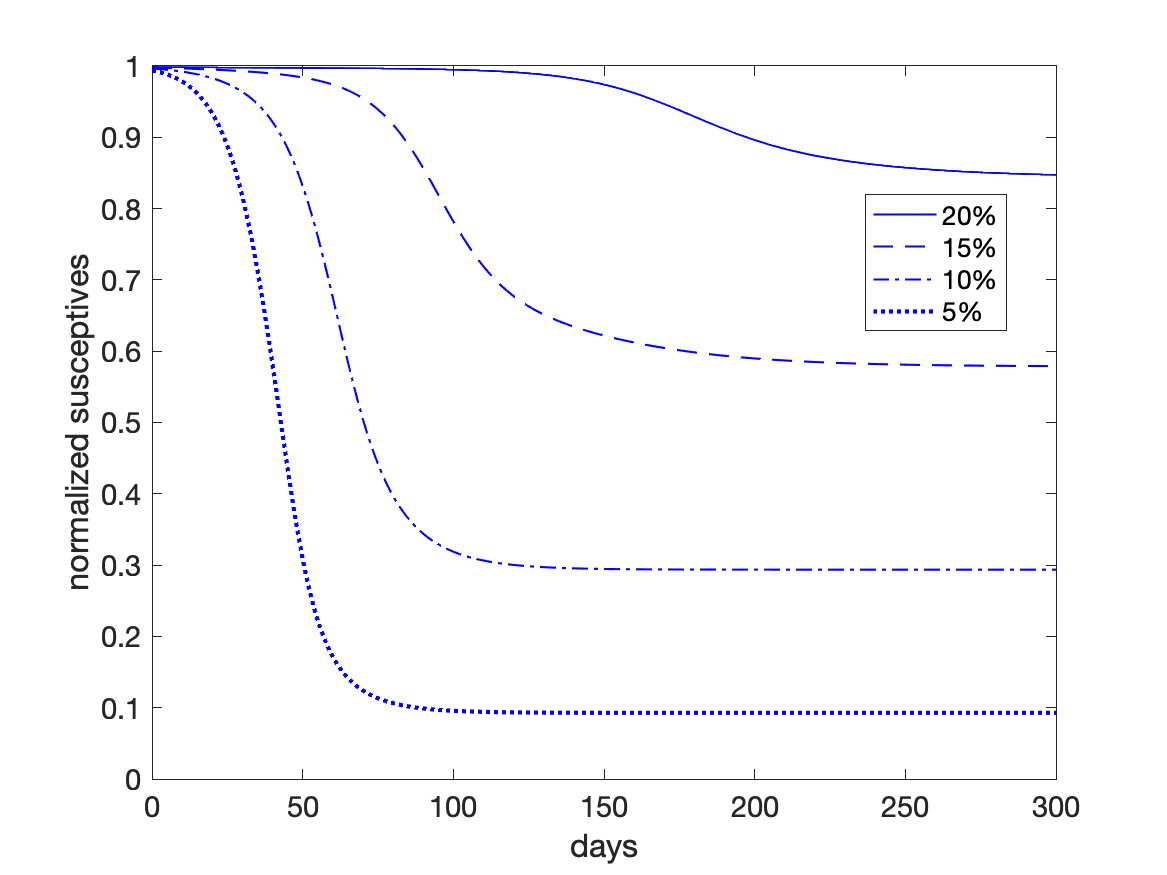}}
\caption{simulated susceptives, using data on 24 May 2020 (day 90) with the specified level of rapid tests.}
\label{fig_test}
\end{figure}

\subsection{Introducing rapid tests for screening}
A way to reduce the impact of the epidemic after the end of the lockdown, could be that of removing infective undiagnosed people by using rapid tests for detecting whether one is infective. Confining the analysis, for sake of simplicity, to the model (\ref{si2r2}), we then introduce an extra-removal term in the second equation, thus getting:

\begin{equation}\label{si3r2}
\left\{ \begin{array}{rcl}
\dot x(t)          &=& -\frac{\beta}N x(t)y_1(t),\\[1mm]
\dot y_1(t)      &=& \frac{\beta}N x(t)y_1(t)-\sigma y_1(t-\tau) s_+(y_1(t))-(\gamma_1+\frac{\nu}N) y_1(t),\\[1mm]
\dot y_2(t)      &=& \sigma y_1(t-\tau)s_+(y_1(t))+\frac{\nu}N y_1(t)-\gamma_2 y_2(t),\\[1mm]
\dot z_1(t)      &=& \gamma_1y_1(t),\\[1mm]
\dot z_2(t)      &=& \gamma_2y_2(t),
\end{array}\right.
\end{equation}
where we have used the same notations as in (\ref{si2r2}), with $\frac{\nu}N$ the fraction of individuals that undergoes the test. Also in this case, one verifies that
the conservation property  (\ref{N}) holds true, with $N$ the total number of individuals. The multi-region variant of (\ref{si3r2}) is analogously derived. 

In Figure~\ref{fig_test} we show the simulation of Italy in case the lockdown is released, by using the parameters estimated at day 90 (as we did before) but neglecting, for sake of simplicity, the migration terms. Consequently, we now allow all the people to become susceptive, though introducing a level of tests of 5\%, 10\%, 15\%, and 20\% in the population. 

As one may see, in order to get a scenario  at day 90 similar to the one yielded by the lockdown (about 15\% of the population exposed to the virus), an amount of rapid tests equal to 20\% of the whole Italian population should be administered on daily basis ($\nu/N=0.2$). The exposed population (susceptives estimated by the model at time $t=0$) would rise to more than 40\%, 70\% and  90\%  if the tested people drop to 15\%, 10\% and 5\% ,respectively. Even though it is not shown in the figure, a level test of 25\% seems to be sufficient to make almost all population not exposed to the virus.  

In conclusion, if rapid tests are administrated randomly to the population, only a high percentage amount   would produce benefits comparable (or even better) to that of imposing a lockdown. Nonetheless, we could expect a significant improvement of their effectiveness, if addressing the rapid test campaign to people subject to a major risk of spreading the epidemic, such as people working in public offices, markets, etc.

\section{Conclusions}\label{fine} In this paper we have shown that the \si2r2 model and, in particular, its multi-region variant mr\si2r2, can be reliably used for predicting the evolution of the COVID-19 epidemic in Italy. It can, therefore, represent a viable tool for optimizing the usage of a healthcare system. We have used it to asses the impact of lockdown on the management of this emergence, as well as to infer that bringing it forward by a couple of weeks would have presumably resulted into a much better outcome of the spread of the epidemic.
We have also derived an estimate for the actual total infected people, i.e., both the diagnosed and the undiagnosed ones, guessing that the former are about 4\% of the total. Last, but not least, the model can be also extended to include the extensive usage of rapid tests to manage the epidemic when the lockdown is released.

\paragraph{Declarations of interest:}  none.

\paragraph{Acknowledgements:} the authors are very indebted to the wonderful team actively contributing the {\em mrSIR project}, including the ``Associazione Italia Digitale'', as well as to the generous supporters \cite{ginger}.

\end{document}